\documentclass[12pt]{elsart}
\usepackage{feynmp,epsfig,graphics}

\def\tr{{\rm tr} \,}

\def\pslash{\FMslash p}

\def\qslash{\FMslash q}

\begin{document}
GSI-Preprint-2002-30
\begin{frontmatter}
\title{Migdal's short range correlations in a covariant model}
\author[GSI]{M.F.M. Lutz $^b$}
\address[GSI]{Gesellschaft f\"ur Schwerionenforschung (GSI),\\
Planck Str. 1, 64291 Darmstadt, Germany}
\address{$^b$Institut f\"ur Kernphysik, TU Darmstadt\\
D-64289 Darmstadt, Germany}
\begin{abstract}
We construct a covariant model for short range correlations of a
pion emerged in nuclear matter. Once the delta-hole contribution
is considered an additional and so far neglected channel opens
that leads to significant modifications in the vicinity of the
kinematical region defined by $\omega \sim |\vec q\,|$. We
speculate that this novel effect should be important for the
quantitative interpretation of charge exchange reactions like
$^{12}$C($^3$He,t).
\end{abstract}
\end{frontmatter}


The pion plays a special role in nuclear physics. A large amount
of work has been done to understand the many facets of nuclear
pion dynamics
\cite{Ericson:Ericson,Oset:Rho,Oset:Weise,Chanfray:Ericson,Dmitriev,Migdal,Delorme:Guichon,Nieves:Oset:Recio}.
Nevertheless, there still exists quite a bit of ambiguity as to
what is the quantitative form of the pion spectral function in
cold nuclear matter. This problem reflects in part the fact that
there are no commonly accepted Migdal parameters that describe the
short range correlation effects. In particular $g'_{N \Delta}$ and
$g'_{\Delta \Delta}$ are not too well determined (see e.g.
\cite{Wambach,Suzuki}). In this short note we focus on the
particular aspect how to treat short range correlations in a
covariant manner. The latter are required to reduce the strength
of the nucleon- and delta-hole contributions to the pion self
energy. This avoids for instance a pion condensate at unreasonably
small nuclear densities. Most works were performed in a
non-relativistic framework with the exception of a series of works
by Dmitriev \cite{Dmitriev,Xia,Dawson,Beck,Dmitriev1993,Nakano}.
In the latter works it was demonstrated that a proper covariant
treatment of the nucleon-hole term leads to a contribution
proportional to $\omega^2 - \vec q\,^2$ rather than to $\vec
q\,^2$ as suggested by a non-relativistic treatment. Thus a
faithful evaluation of the pion self energy in nuclear matter
requires a relativistic treatment at least in the vicinity of
$\omega \sim |\vec q\,|$. The purpose of this work is to
demonstrate that previous works
\cite{Dmitriev,Xia,Dawson,Beck,Dmitriev1993,Nakano} did not
completely succeed in constructing a covariant model for Migdal's
short range correlations. The shortcomings of these works will be
overcome and a fully covariant generalization of Dmitriev's model
will be presented here. It turns out that the necessary
modifications to Dmitriev's model lead to significant effects in
the pion self energy for $\omega \sim |\vec q\,|$ missed so far.

Following Dmitriev's original work \cite{Dmitriev} we consider the
interaction of pions with nucleons and isobars in terms of the
leading order vertices
\begin{eqnarray}
{\mathcal L}&=& \frac{f_{N}}{m_\pi}\,\bar \psi
\,\gamma_5\,\gamma^\mu\,\vec \tau \,\partial_\mu \vec \phi_\pi
\,\psi+ \frac{f_{\Delta }}{m_\pi} \,\Big( \bar \psi^\mu \,\vec T
\,\psi\, \partial_\mu \vec \phi_\pi + {\rm h.c} \Big)\,,
\label{pin-pid}
\end{eqnarray}
as predicted by the chiral Lagrangian. We use $T^\dagger_i\,T_j =
\delta_{ij}- \tau_i\,\tau_j /3$ and $f_{N}=0.988$ and $f_{\Delta
}= 2\,f_{N}$ in this work. The nucleon and  isobar propagators are
\begin{eqnarray}
&& S (p,u) = \frac{1}{\pslash -m_N + i\,\epsilon} + \Delta S(p,u)
\,, \quad \nonumber\\
&& \Delta S (p,u) = 2\,\pi\,i\,\Theta \Big(p\cdot u \Big)\,
\delta(p^2-m_N^2)\,\big( \pslash +m_N \big)\,\Theta
\big(k_F^2+p^2-(u\cdot p)^2\big)\,,
\nonumber\\
&& S_{\mu \nu}(p) =\frac{-i}{\pslash-m_\Delta+ i\,\epsilon} \left(
g_{\mu \nu}-\frac{\gamma_\mu\,\gamma_\nu }{3} -\frac{2\,p_\mu
\,p_\nu }{3\,m_\Delta^2} -\frac{\gamma_\mu \,p_\nu -p_\mu\,
\gamma_\nu}{3\,m_\Delta} \right) \,,\label{iso-prop}
\end{eqnarray}
where the time like four-vector $u_\mu$ specifies the nuclear
matter frame. For symmetric nuclear matter at rest it follows
$u_\mu = (1, \vec 0 \,)$ and $\rho = 2\,k_F^3/(3\pi^2)$. In
(\ref{iso-prop}) we do not consider an additional structure in the
isobar propagator which would modify the off-shell properties of
the isobar only. This term is not relevant for the development of
this work.

It is straight forward to write down the nucleon and delta-hole
contributions to the pion self energy,
\begin{eqnarray}
\Pi_\pi (q,u) = -q^\mu\, \Big(\Pi_{\mu \nu}^{(Nh)}(q,u)+\Pi_{\mu
\nu}^{(\Delta h)}(q,u) \Big)\,q^\nu  \,, \label{}
\end{eqnarray}
where we follow previous works and disregard vacuum polarization
effects. This leads to loop functions of the form
\begin{eqnarray}
&& \! \Pi_{\mu \nu}^{(\Delta h)}(q,u) = \frac{4}{3}\,
\frac{f^2_{\Delta }}{m^2_\pi} \int \frac{d^4 p}{(2\pi)^4}\,i\,\tr
\,\Delta S(p,u)\,  S_{\mu \nu }(p+q) + (q_\mu
\to -q_\mu )  \,, \nonumber\\
&& \!  \Pi_{\mu \nu}^{(Nh)}(q,u) = 2\,\frac{f^2_{N}}{m^2_\pi}
\,\int \frac{d^4 p}{(2\pi)^4}\,i\,\tr \,\Bigg( \Delta S(p,u)\,
\gamma_5 \,\gamma_\mu \,\frac{1}{\pslash +\qslash -m_N}
 \,\gamma_5 \,\gamma_\nu \nonumber\\
&& \qquad \qquad \qquad \quad + \frac{1}{2}\, \Delta S(p,u)\,
\gamma_5 \,\gamma_\mu \,\Delta S(p+q) \,\gamma_5 \,\gamma_\nu
\Bigg)
 + (q_\mu
\to -q_\mu ) \,.  \label{nh-dh}
\end{eqnarray}
In order to arrive at realistic nucleon-hole and delta-hole
contributions to the pion polarization in nuclear matter it is
crucial to introduce short range correlation effects that
significantly reduce the contributions of the nucleon-hole and
delta-hole diagrams \cite{Migdal}. Most authors would also argue
that the loop functions (\ref{nh-dh}) should be multiplied by a
suitable form factor that reduces the strength of the loop
function at large momenta. Since this would necessarily introduce
some ambiguities we refrain from doing this here. The aim of this
short note is not to provide a fully realistic pion self energy in
nuclear matter rather we suggest to improve a frequently used
model as to achieve consistency with covariance. We do, however,
incorporate a reasonable spectral distribution of the isobar and
fold the delta-hole loop function with a spectral function that
describes the $P_{33}$ phase shift of pion-nucleon scattering. The
specifics of the spectral function we use here can be found in
\cite{LWF} \footnote{There is a misprint in (131) of \cite{LWF}.
Replace $C^2 \to m_\Delta \,C^2$ and use $C=1.85$.}.

A covariant from of the short range correlations was introduced
explicitly by Nakano et al. \cite{Nakano}
\begin{eqnarray}
{\mathcal L}_{\rm Migdal} &=&g_{11}'\, \frac{f^2_{N}}{m^2_\pi} \,
\Big(\bar  \psi \,\gamma_5\,\gamma_\mu\,\vec \tau \,\psi \Big)
\,\Big(\bar  \psi \,\gamma_5 \,\gamma^\mu\,\vec \tau \,\psi\Big)
\nonumber\\
&+& g_{22}'\, \frac{f^2_{\Delta }}{m^2_\pi} \, \Bigg( \Big(\bar
\psi_\mu \,\vec T \,\psi \Big) \,\Big(\bar \psi \,\vec T
\,\psi^\mu\Big) + \left(\Big(\bar \psi_\mu \,\vec T \,\psi \Big)
\,\Big(\bar \psi^\mu \,\vec T \,\psi\Big)+{\rm h.c.}\right) \Bigg)
\nonumber\\
&+& g_{12}'\, \frac{f_{N}\,f_\Delta}{m^2_\pi} \,\Big(\bar \psi
\,\gamma_5\,\gamma_\mu\,\vec \tau \,\psi \Big) \left( \Big(\bar
\psi^\mu \, \vec T \,\psi\Big) +{\rm h.c.}\right)
\,,\label{cov-Migdal}
\end{eqnarray}
where it is understood that the local vertices are to be used at
the Hartree level. The Fock contribution can be cast into the form
of a Hartee contribution by a simple Fierz transformation.
Therefore it only renormalizes the coupling strength in
(\ref{cov-Migdal}) and can be omitted here. Note that the terms
proportional to $\bar \psi_\mu\,\bar \psi^\mu$ and $
\psi_\mu\,\psi^\mu$ of (\ref{cov-Migdal}) were missed in
\cite{Nakano}. They are required to recover the proper non
relativistic limit of the short range correlations as introduced
by Migdal.

In previous works
\cite{Dmitriev,Xia,Dawson,Beck,Dmitriev1993,Nakano} the short
range correlation effects were introduced in the form,
\begin{eqnarray}
&& \!\frac{\Pi (q,u)}{q^2} = \frac{\Pi_{Nh}\,\big(q^2+
g_{22}'\,\Pi_{\Delta h}\big)+ \Pi_{\Delta
h}\,\big(q^2+g_{11}'\,\Pi_{Nh}\big) -2\,g_{12}\,\Pi_{Nh}\,
\Pi_{\Delta h}}{\big(
q^2+g_{11}'\,\Pi_{Nh}\big)\,\big(q^2+g_{22}'\,\Pi_{\Delta
h}\big)-g_{12}'^2\,\Pi_{N h}\, \Pi_{\Delta h}} \,,
\nonumber\\
&&  \Pi_{\Delta h}(q,u) =-q^\mu\, \Pi_{\mu \nu}^{(\Delta
h)}(q,u)\,q^\nu \,, \qquad  \Pi_{Nh}(q,u) =-q^\mu\, \Pi_{\mu
\nu}^{(Nh)}(q,u)\,q^\nu \,, \label{app-sr}
\end{eqnarray}
with  $\Pi_{\mu \nu}^{(N h)}$ and $  \Pi_{\mu \nu}^{(\Delta h)} $
taken  from (\ref{nh-dh}). As will be demonstrated explicitly
below the result (\ref{app-sr}) is strictly speaking not correct
and requires a generalization for $$q^2 \neq (q\cdot u)^2\,.$$ The
form of (\ref{app-sr}) was taken over from a corresponding
expression obtained by Migdal \cite{Migdal} in the
non-relativistic case, but not properly modified for the covariant
vertices (\ref{pin-pid}, \ref{cov-Migdal}).

It is evident that the expression (\ref{app-sr}) were correct if
the generic loop functions $\Pi_{\mu \nu}^{(N h)}$ and $\Pi_{\mu
\nu}^{(\Delta h)}$ had contributions proportional to $g_{\mu \nu}$
and $q_\mu \,q_\nu$ only. However, the most general decomposition
of the loop functions involves additional structures proportional
to $u_\mu\,q_\nu$ and $q_\mu\,u_\nu $. To derive the correct
generalization of (\ref{app-sr}) we introduce a transverse
projector, $T^{\mu \nu}(q,u)$ and a set of longitudinal projectors
$L_{ij}^{\mu \nu}(q,u)$,
\begin{eqnarray}
&& T_{\mu \nu} = g_{\mu \nu}-\frac{q_\mu\,q_\nu}{q^2}-X_\mu\,X_\nu
\,,\qquad X_\mu = \frac{(q\cdot u )q_\mu-
q^2\,u_\mu}{\sqrt{q^2}\,\sqrt{q^2-(q\cdot u)^2}} \,,
\nonumber\\
 && L^{\mu \nu}_{11} = \frac{q^\mu\,q^\nu}{q^2} \,,
\quad L_{12}^{\mu \nu }= \frac{q^\mu\,X^\nu}{\sqrt{q^2}}\,,  \quad
L_{21}^{\mu \nu }= \frac{X^\mu\,q^\nu}{\sqrt{q^2}}\,,\quad
 L_{22}^{\mu \nu} = X^\mu \,X^\nu\,, \nonumber\\
&&  L_{ik}\,\cdot L_{lj} = \delta_{kl}\,L_{ij}\,, \qquad L_{ij}
\cdot T =0 = T \cdot L_{ij} \,, \quad u^2=1 \,,\quad X^2=1
\,,\label{add22a}
\end{eqnarray}
that trivialize the solution of the Dyson equation in the presence
of the structures $q_\mu\,u_\nu$ and $u_\mu\,q_\mu $. The loop
functions are decomposed into the complete set of projectors,
\begin{eqnarray}
\Pi_{ij}^{(N h)} = \Pi_{\mu \nu}^{(N h)}\,L^{\mu \nu}_{ij} \,,
\qquad  \Pi_{ij}^{(\Delta  h)} = \Pi_{\mu \nu}^{(\Delta
h)}\,L^{\mu \nu}_{ij} \,, \label{add22b}
\end{eqnarray}
where only the longitudinal projections are needed in this work. A
corresponding decomposition of the Migdal interaction vertices,
$$
\gamma_5 \,\gamma^\mu \otimes \gamma_5\,\gamma_\mu = \gamma_5
\,\gamma_\mu \otimes \gamma_5\,\gamma_\nu \,\Big(T^{\mu \nu}+
L_{11}^{\mu \nu}+ L_{22}^{\mu \nu} \Big)\,,
$$
implies a straight forward generalization of (\ref{app-sr}). The
self energy can be cast into the form of a sum of $11$, $33$ and
$13$, $31$ components of an appropriate 4$\times$4 matrix that
incorporates the additional 2$\times$2 matrix structure introduced
in (\ref{add22a},\ref{add22b}),
\begin{eqnarray}
\frac{\Pi}{q^2} &=& -\Big[\Big( 1-J\,G \Big)^{-1} \,J \Big]_{11} -
\Big[\Big( 1-J \,G\Big)^{-1} \,J  \Big]_{33}
\nonumber\\
&-& \Big[\Big( 1-J \,G\Big)^{-1} \,J \Big]_{13} - \Big[\Big(
1-J\,G \Big)^{-1}\,J  \Big]_{31} \,, \label{full-result}
\end{eqnarray}
where the coupling matrix, $G$, and loop matrix, $J$, are
\begin{eqnarray}
&& G = \left(
\begin{array}{llll}
g_{11}' & 0 & g_{12}' & 0 \\
0 & g_{11}' & 0 & g_{12}'  \\
g_{12}' & 0 & g_{22}' & 0 \\
0 & g_{12}' & 0 & g_{22}'
\end{array}
\right) \,, \qquad J = \left(
\begin{array}{llll}
\Pi_{11}^{(N h)} & \Pi_{12}^{(N h)} & 0 & 0 \\
\Pi_{21}^{(N h)} & \Pi_{22}^{(N h)} & 0 & 0  \\
0 & 0 & \Pi_{11}^{(\Delta h)} & \Pi_{12}^{(\Delta h)} \\
0 & 0 & \Pi_{21}^{(\Delta h)} & \Pi_{22}^{(\Delta h)}
\end{array}
\right)\,. \label{def-matrix}
\end{eqnarray}
It is evident that only matrix elements $ij$ with $i,j=1,3$
contribute to the self energy since the latter reflect the
derivative coupling of the pion baryon vertex structure.  In the
particular case where $\Pi_{12}^{(N h)} = \Pi_{21}^{(N h)}=0 $ and
$\Pi_{12}^{(\Delta h)} = \Pi_{21}^{(\Delta h)}=0 $ holds the
result (\ref{full-result}) reproduces (\ref{app-sr}).

We proceed and derive explicit representations for the loop matrix
$J$ for the standard case of nuclear matter at rest with $u_\mu
=(1, \vec 0\,)$. The nucleon-hole functions read\footnote{The
off-diagonal (diagonal) loop function $\Pi_{ij}(\omega, \vec
q\,)=\Pi^{(\Delta h)}_{12}(-\omega, \vec q\,)$ must be
anti-symmetrized (symmetrized) in $\omega$. This follows from the
corresponding property of $L^{\mu \nu}_{ij}$. The resulting pion
self energy is symmetric in $\omega$.},
\begin{eqnarray}
\Pi^{(Nh)}_{ij}(\omega, \vec q\,) &=& \frac{i\,f^2_{N}}{m^2_\pi}
\Im
\int_0^{k_F}\,\frac{d^3p}{2\,p_0\,(2\pi)^3}\,\frac{8\,K_{ij}^{(Nh)}\,\Theta
\Big( |\vec p+\vec q\,|-k_F\Big)}{ 2\,p\cdot q+q^2+i\,\epsilon
}\,\,\Theta \big(p_0+\omega \big)
\nonumber\\
&+& \frac{f^2_{N}}{m^2_\pi}\, {\mathcal P }
\int_0^{k_F}\,\frac{d^3p}{2\,p_0\,(2\pi)^3}\,\frac{8\,K_{ij}^{(Nh)}}{
2\,p\cdot q+q^2+i\,\epsilon}
 \pm (q_\mu \to -q_\mu )\,, \label{matrix-nh-a}
\end{eqnarray}
where $q_\mu =(\omega , \vec q\,)$, $p_0 = \sqrt{m_N^2+\vec
p\,^2}$ and
\begin{eqnarray}
&& K_{11}^{(Nh)} = 2\,m_N^2 \,,\qquad
K_{12}^{(Nh)}=K_{21}^{(Nh)}=0 \,, \qquad
\nonumber\\
&& K^{(Nh)}_{22}= \frac{\omega^2 -\vec q\,^2}{\vec q\,^2}\,\Big(
2\, \vec p\,^2 + \omega \, p_0 + \vec p \cdot \vec q \Big)\,.
\label{matrix-nh-b}
\end{eqnarray}
Since the transition loop functions $\Pi_{12}^{(Nh)} =
\Pi_{21}^{(Nh)} =0$ vanish identically  it would be justified to
ignore the coupled channel structure in the nucleon-hole channel
if the delta-hole contributions were neglected all together. The
loop function $\Pi_{22}^{(Nh)} \neq 0$ will be relevant once the
delta-hole channel is considered since in the latter channel
$\Pi_{12}^{(\Delta h)} \neq 0 $ holds. The following result is
derived,
\begin{eqnarray}
&& \Pi^{(\Delta h)}_{ij}(\omega, \vec q\,) = \frac{4}{9}
\frac{f^2_{\Delta }}{m^2_\pi}
\int_0^{k_F}\,\frac{d^3p}{2\,p_0\,(2\pi)^3}\,\frac{8\,K_{ij}^{(\Delta
h)}\,\big(m_N\,m_\Delta +m_N^2+(p\cdot q )\big)}{ 2\,p\cdot
q+q^2-m_\Delta^2+m_N^2+i\,\epsilon } \nonumber\\
&& \qquad \qquad \quad \;\pm (q_\mu \to -q_\mu )\,, \nonumber\\
\nonumber\\ && K_{11}^{\Delta h} = 1- \frac{(q^2+p\cdot
q)^2}{q^2\,m^2_\Delta} \,,\quad K_{22}^{\Delta h} = 1 +
\frac{(\omega \,|\vec p \,|\,\cos
(\vec q\,,\vec p\,) - |\vec q\,|\,p_0 )^2}{m_\Delta^2\,q^2} \,, \nonumber\\
&& K_{12}^{\Delta h} = K_{21}^{\Delta h} = i\,\frac{q^2+p\cdot q
}{q^2\,m_\Delta^2} \,\big( |\vec q\,|\,p_0-\omega \,|\vec p
\,|\,\cos (\vec q\,,\vec p\,) \big) \,. \label{matrix-dh}
\end{eqnarray}
We refrain from presenting analytic expressions for the final loop
functions since the expressions are cumbersome and already
published for the 11 components \cite{Beck,Nakano}. Moreover it is
more transparent to work directly with the representations
(\ref{matrix-nh-a}, \ref{matrix-nh-b}, \ref{matrix-dh}).

\begin{figure}[t]
\epsfysize=10cm
\begin{center}
\mbox{\epsfbox{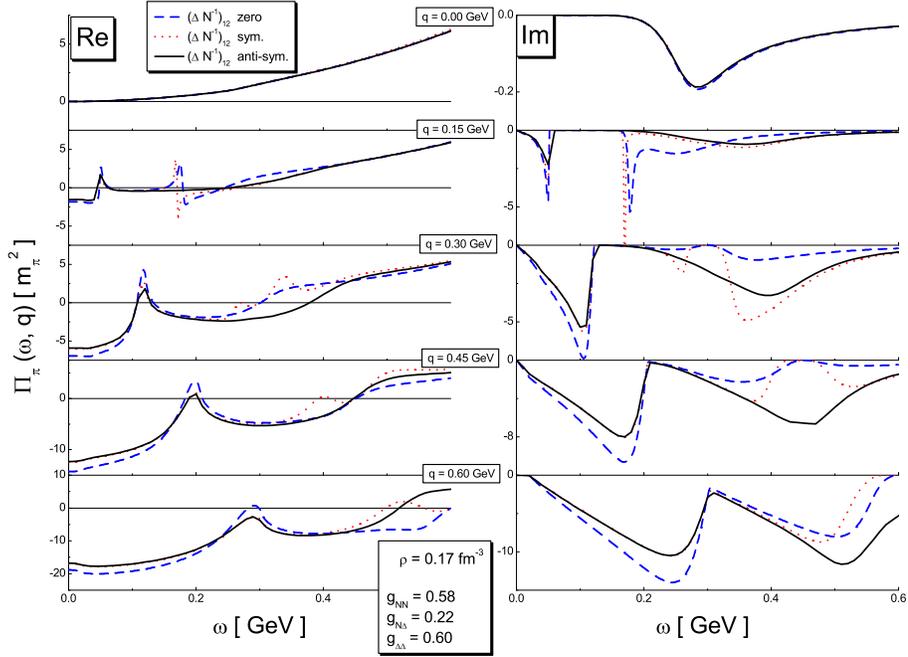}}
\end{center}
\caption{The pion self energy as a function of $\omega$ and $|\vec
q |$ evaluated at nuclear saturation density with $k_F=268$ MeV.
The solid line shows the full result including the coupled channel
structure of (\ref{full-result}), the dashed line follows with
$\Pi_{12}^{(\Delta h)}\to 0 $ in (\ref{full-result}). The dotted
line gives the result if one symmetrized the transition
isobar-hole loop function $\Pi_{12}^{(\Delta h)}$.} \label{fig1}
\end{figure}

In Fig. 1) we present the pion self energy (\ref{full-result}) and
(\ref{app-sr}) for the choice of parameters $g_{11} = 0.585$ and
$g_{12} = 0.191 + 0.051\,g_{22}$ with $g_{22} = 0.6$ as suggested
in \cite{Oset:Rho} and \cite{Nakano}. We do not study here
possible deviations from those values because this is not the
point of this work. The figure clearly illustrates significant
effects in certain kinematical regions as the result of a proper
treatment of the coupled channels. Most striking is the
enhancement by about a factor 5 found for the imaginary part of
the pion self energy at $\omega \sim$ 350 MeV and $|\vec q\,| =
300$ MeV. The inclusion of  the transition loop function
$\Pi_{12}^{(\Delta h )} $ together with $\Pi_{22}^{(\Delta h )}$
is crucial here. From the form of $\Pi_{12}^{(\Delta h )} $ it
follows directly that at $\omega \neq 0$ and $|\vec q\,| = 0$, the
kinematical region probed by the quenching of the Gamow-Teller
resonance \cite{Bang}, the coupled channel structure discussed
here is superficial. In this case the use of the expression
(\ref{app-sr}) is justified since $\Pi_{12}^{(\Delta h )} = 0$
holds. A further interesting limit is $\omega =0$ with $\vec
q\,\neq 0$ as is probed by a possible pion condensate. Here one
would expect to recover the algebraic form of the non-relativistic
scheme of Migdal for $|\vec q \,| \ll m_N $ and $|\vec p \,| \ll
m_N$. This is indeed confirmed if the functions $K_{ij}^{(\Delta
h)}$ are expanded in powers of $$ \frac{ | \vec p\,|}{m_{N}} \,,
\qquad \,
 \frac{ | \vec q\,|}{m_{N}}.$$
Then the transition moment, $K_{12}^{(\Delta h)}$, is suppressed
by the factor $| \vec q\,|/m_{N}$ as compared to the leading
moment, $K_{11}^{(\Delta h)}$. Therefore we expect only minor
corrections when studying pion condensation phenomena within the
generalized result (\ref{cov-Migdal}). Particularly striking are,
however, new effects close to the kinematical point $\omega^2 \sim
\vec q\,^2 \neq 0$ where the transition loop function can no
longer be neglected. This kinematical domain is crucial for the
peak structure \cite{Dmitriev,Delorme:Guichon,Dmitriev1993} in the
$^{12}$C($^3$He,t) transfer reaction \cite{Gaarde,Roy-Stephan}.

{\bfseries{Acknowledgments}}

I would like to thank E. Kolomeitsev, C. Korpa and  D.
Voskresenski for useful discussions.


\end{document}